\documentclass[twocolumn,nofootinbib,aps,preprintnumbers,amsmath,amssymb]{revtex4-1}
\usepackage{amsmath}
\usepackage{graphicx}
\usepackage{dcolumn}
\usepackage{bm}
\usepackage{epsfig,color,xspace,multirow,xr,bbold}
\usepackage[all]{xy}
\usepackage{setspace}
\usepackage{url}

\newcommand{\br}{{\bf r}}
\newcommand{\rd}{{\rm d}}
\newcommand{\re}{{\rm e}}

\begin{document}
\title{Transferable atomic multipole machine learning models for small organic molecules}

\author{Tristan Bereau}
\email{bereau@mpip-mainz.mpg.de}
\author{Denis Andrienko}
\affiliation{Max Planck Institute for Polymer Research, Ackermannweg 10, 55128
  Mainz, Germany}
\author{O. Anatole von Lilienfeld}
\email{anatole.vonlilienfeld@unibas.ch}
\affiliation{Institute of Physical Chemistry and National Center for Computational
Design and Discovery of Novel 
Materials, Department of Chemistry,
  University of Basel, 4056 Basel, Switzerland} 
\affiliation{Argonne Leadership Computing Facility, Argonne National
  Laboratory, Argonne, Illinois 60439, USA}
\date{\today}

\begin{abstract}
 Accurate representation of the molecular electrostatic potential, which is often expanded in distributed multipole moments, is crucial for an efficient evaluation of intermolecular interactions. Here we introduce a machine learning model for multipole coefficients of atom types H, C, O, N, S, F, and Cl in any molecular conformation. The model is trained on quantum chemical results for atoms in varying chemical environments drawn from thousands of organic molecules. Multipoles in systems with neutral, cationic, and anionic molecular charge states are treated with individual models. The models' predictive accuracy and applicability are illustrated by evaluating intermolecular interaction energies of nearly 1,000 dimers and the cohesive energy of the benzene crystal.
\end{abstract}
\maketitle

\section{Introduction}

Efficient evaluation of intermolecular (also termed as van der Waals~\cite{AtkinsPC,IsraelachviliForces,WedlerPC}) interactions is an essential part of all classical molecular dynamics simulations. Electrostatic, induction, dispersion, and exchange-repulsion are the most frequently encountered non-bonded contributions to the energy of interaction between molecules. In order to boost computational efficiency, these contributions are often projected on pairwise-additive functions, the sum of which then approximates the potential energy surface of a molecular assembly. Many-body effects (e.g., induction and dispersion) are accounted for effectively, by an appropriate parametrization of the potential energy surface. These parametrizatons are, by construction, state-point dependent and rely on either measured or first-principles evaluated thermodynamic properties of a molecular assembly at a certain state point. For example, partial charges and Lennard-Jones parameters are often adjusted to fit the density, heat of vaporization, and other thermodynamic properties~\cite{jorgensen1996development}. 

Force field transferability and accuracy can of course be improved by retaining many-body contributions. The decisive advantage of this approach, which justifies extra computational effort, is that these terms can be evaluated perturbatively, i.e., by first calculating electronic properties of non-iteracting molecules using first-principles methods and then accounting for electrostatic (first order), induction (second order), and dispersion (higher orders) contributions in a perturbative way~\cite{stone2013theory}. Such parametrizations do not require experimental input, are state-point independent and, as such, can be used to pre-screen chemical compounds {\em in silico}.

In this approach, however, even the molecular electrostatic potential must be evaluated for every single molecular conformation, requiring  electronic structure calculations at practically every molecular dynamics step. It has also been pointed out that the multipole-moment (MTP) description of the electrostatics must include not only atomic charges but also higher moments (e.g., dipoles and quadrupoles)~\cite{kramer2014charge, piquemal2007toward,gresh2007anisotropic,gordon2013accurate}, improving free-energy calculations~\cite{ponder2010current, bereau2013leveraging}, spectroscopic signatures~\cite{lee20132d,cazade2014computational}, and dynamics~\cite{jakobsen2015multipolar}.

Avoiding the need for frequent quantum-chemical calculations has motivated the development of fast prediction methods, such as machine learning (ML)~\cite{cortes1995support,muller2001introduction,scholkopf2002learning}. With ML we refer to statistical algorithms that extract correlations 
by training on input/output data, and that improve in predictive power as more training data is added~\cite{witten2005data}. While ML models for the fitting of potential energies have been in use for decades~\cite{SumpterNoidNeuralNetworks1992}, the possibility to infer point charges, MTPs and polarizabilities has been investigated only recently~\cite{rai2013fast, ivanov2015genetic, handley2009optimal, mills2012polarisable}. These approaches interpolate between a large number of conformations to accurately describe the effects of changes in the geometry.  The accuracy that is reached comes at a price: The specificity of the learning procedure limits its applicability to the given molecule of interest. Instead of training electrostatic models for every new molecule, here we construct a {\em transferable} MTP model which can be applied not only to different molecular conformers but also atom types. 

The paper is organized as follows. We first describe how to build a machine learning MTP model that predicts static, atomic point charges, dipoles, and quadrupoles for H, C, O, N, S, F, and Cl atom types in specific chemical environments. Next, the resulting electrostatic interactions are combined with a classical many-body dispersion (MBD)~\cite{bereau2014toward} in order to validate the model by estimating intermolecular energies of nearly 1,000 molecular dimers as well as the cohesive binding energy of the benzene crystal. We find that the machine-learning model retains an accuracy similar to the same model parametrized from individual quantum-chemical calculations.






\section{Methods}

The following describes the ML model, the baseline property used in the delta-learning
procedure, the dataset, and the description of the reference MTPs. 

\subsection{Machine Learning model}

We rely on supervised learning to construct a kernel-ridge regression
which generalizes the linear ridge regression model (i.e., linear
regression with regularizer $\lambda$) by mapping the input space $x$ into a
higher-dimensional ``feature space,'' $\phi(x)$, thereby casting the problem
in a linear way~\cite{MueMikRaeTsuSch01,HasTibFri01}.
The strength of the method comes from avoiding the actual determination 
of $\phi$ thanks to the so-called kernel trick \cite{scholkopf1998nonlinear}:
Since the ML algorithm only requires the inner product between data vectors in
feature space, one can apply a kernel function $k(x,x')$ to compute dot products
within input space, thereby leaving the feature space entirely implicit.
As a result, the problem is reformulated from a $v$-dimensional input space
(i.e., the dimensionality of each data vector) into an $n$-dimensional space
spanned by the number of samples in the training set.  This characteristics
implies that the larger $n$, the better the prediction ought to be---thus the
denomination of a \emph{supervised learning} method.

Here, we build on the $\Delta$-ML approach~\cite{DeltaPaper2015}, which estimates the difference between the desired property
and an inexpensive baseline model that accounts for the most relevant physics. 
More specifically, a refined target property $p(x)$ is predicted
from baseline property $p^{\rm Vor}$ (see Sec.~\ref{sec:vor}) plus the ML-model $\Delta$
\begin{equation}
  p(x) = p^{\rm Vor}(x) + \Delta(x,p^{\rm Vor}),
\end{equation}
where $x$ corresponds to the representation vector---or~descriptor---of the
input sample (e.g., query molecule). 
$\Delta$ corresponds to the standard kernel-ridge regression model of the difference
between baseline and target property constructed for $n$ training samples,
\begin{equation}
  \Delta(x,p^{\rm Vor}) = \sum_{i=1}^n \alpha_i \left[ k(x,x_i)+ k'(p^{\rm Vor},p^{\rm Vor}_i) \right],
  \label{eq:MLmodel}
\end{equation}
where $\alpha_i$ is the weight given to training molecule $i$. 
These weights are determined by best 
reproducing the reference property
$p^{\rm Ref}(x)$ for each sample in the training set
according to the closed-form solution ${\bm \alpha} = \left({\bf K} + {\bf K}' +
\lambda \mathbb{1}\right)^{-1}({\bf p}^{\rm Ref}-{\bf p}^{\rm Vor})$, where 
${\bf p}^{\rm Ref}-{\bf
  p}^{\rm Vor}$ is the vector of training properties, 
  i.e. difference between reference and
baseline, and ${\bf K}$ and ${\bf K}'$ are the two kernel matrices.  
Note that in Eq.~\ref{eq:MLmodel}, 
we have included representation {\em and} baseline property in the kernel,
each having a different width in their respective kernel functions. 

ML maps an input representation vector $x$ into a scalar value of similarity. 
Thus, before applying ML to predict atomic MTPs, the information contained in the
three-dimensional structure of a molecule must be encoded in a vector of
numbers i.e., its representation or descriptor.  Ideally, this information should reflect symmetries of molecular structures with respect to rotations, translations, reflections, atom index permutations, etc.  Here, we rely on the Coulomb matrix descriptor~\cite{rupp2012fast},
\begin{equation}
  C_{ij} = \begin{cases}
\frac 12 Z_i^{2.4} & \forall\ i=j \\
\frac{Z_iZ_j}{|{\bf R}_i-{\bf R}_j|} & \forall\ i\ne j, \\
\end{cases} 
\end{equation}
where $i$ and $j$ index atoms in the molecule, $Z_i$ is atom $i$'s atomic
number, and ${\bf R}_i$ represents its Cartesian coordinates. 
Note that the Coulomb matrix not only encodes inverse pairwise distances 
between atoms but also the chemical elements involved. 
As different molecules have different numbers of atoms, their Coulomb
matrices will vary in size.  Distant neighbors are 
expected to have a comparatively small impact on a prediction, such that
the inclusion of all neighbors can prove inefficient for large molecules. 
Given a set of molecules, we pad matrices with
zeros such that their size amounts to $n \times n$, where $n$ is the number of
closest neighboring atoms considered \cite{rupp2012fast}.  
In the following, we set $n=4$. 
Given a molecule's $d$ atoms, 
there are $d$ individual atomic MTP samples for the ML to learn from. 
For each, an individual Coulomb matrix is built 
in which the atom of interest fills up the first row/column, 
while the indices of the surrounding $n$ atoms are sorted 
according to the atoms' Euclidean distances to the query atom.
As such, we coarsen our descriptor to contain at least the first
shell of $n$ covalently bound neighbors, and atoms that only
differ in their environment at larger distances will be assigned
the same MTP. We have found $n=4$ to correspond to a reasonable compromise
between computational efficiency and performance.
Note, however, that while such choices of descriptor typically do 
affect the model's performance for given training sets,
other descriptor choices could work just as well---as long as
they meet the requirements and invariances necessary for the
ML of quantum properties~\cite{FourierDescriptor}.

In the context of applying ML to the prediction of tensorial quantities,
such as MTPs, properties $p^{\rm Vor}(x)$ and $p(x)$ will be expressed as
\emph{vectors} of size $m$---the number of independent coefficients of the
tensor of interest (e.g., 1 for a scalar charge, 3 for a vector dipole 
moment, 5 for a traceless second-rank tensor quadrupole). 
We express MTP moments with their minimal number of
independent coefficients by using the spherical-coordinate representation.  
We recognize that the kernel matrices, ${\bf K}$ and ${\bf  K}'$, will remain unmodified when learning/predicting different tensor components 
of the same input data vector.  
Finally, the weights $\alpha$ are expressed as a matrix of 
size $m \times n$, which naturally reduces to a vector
when predicting a scalar quantity.

For this work, we have used the Laplacian kernels,
\begin{align}
	\label{eq:laps}
  k(x_i,x_j) &= \exp\left( -\frac {\left|x_i-x_j\right|}{\sigma N_t} \right), \\
  \label{eq:lapz}
  k'(p^{\rm Vor}_i,p^{\rm Vor}_j) &= \exp\left( -\frac {\left|p^{\rm Vor}_i-p^{\rm Vor}_j\right|}{\zeta N_t}
   \right), 
\end{align}
where $\sigma$ and $\zeta$ are free parameters, $|\cdot|$ corresponds to the
Manhattan, or city block, $L_1$ norm. 
This combination of kernel functions and distance measure has previously been shown to yield the best performance for the modeling of molecular atomization energies and other electronic properties using the Coulomb-matrix representation~\cite{AssessmentMLJCTC2013,singlekernel2015}.
$N_t$ is the number of occurrences of   
the chemical element type to which atom $i$ belongs.
As a result $N_t$ normalizes the width to be consistent
with training set size of a given chemical element. 
We report below (Table II) the strong variance of occurrence
numbers of chemical elements in the employed training set. 
Hyperparameter optimization on 85\% of the elements encompassing 
the training set (see below) yielded $\sigma = 0.005$, $\zeta = 0.002$, and
$\lambda = 10^{-9}$. 
We have subsequently used these parameters for element-specific models throughout.  
Combining
Eqs.~\ref{eq:MLmodel}, \ref{eq:laps}, and \ref{eq:lapz}, 
the $\Delta$-learning ML model predicts the deviation from the 
Voronoi baseline for a new query atom $x$ of element $t$ with Voronoi property
$p^{\rm Vor}$ according to,
\begin{equation}
  \Delta(x,p^{\rm Vor}) = \sum_{i=1}^n \alpha_i \left( \re^{-\frac{|x-x_i|}
  {\sigma N_t}} +
    \re^{-\frac{|p^{\rm Vor}-p^{\rm Vor}_i|}{\zeta N_t }} \right).
\end{equation}

For the modeling of MTP in positively and negatively charged molecules ($\pm 1~e$), 
we have trained respectively different ML models for the same set of molecules, 
systematically assigning the corresponding global molecular charge and assuming a doublet
state.

\subsection{Multipole moments}
\label{sec:rot}
Molecular electron densities were computed using density functional theory calculations at
the M06-2X level of theory~\cite{zhao2008exploring} and cc-pVDZ basis set~\cite{dunning1989gaussian}.  
All {\em ab-initio} calculations were performed using the Gaussian09 program \cite{frisch2009gaussian}.  

The Generalized Distributed Multipole Analysis (GDMA) \cite{stone2013theory} allowed us to partition 
the density into atomic MTPs up to quadrupoles, where we relied on grid-based
quadrature (i.e., switch value of 4).  The same protocol was applied to train the ML models for positively and negatively charged molecules,
after reassigning the global charge of each molecule.

The reference MTPs, obtained from the distributed multipole analysis were rotated into a molecular reference frame, which was constructed from the (sorted) eigenvectors of the molecule's moment of inertia tensor centered at the atom in question.  To ensure uniqueness, we set the positive axis of each vector such that its scalar product with the vector pointing from the atom of interest to the molecule's center of mass is positive. 
For
linear (e.g., diatomic) molecules, we assign the interatomic direction as the first axis and arbitrarily construct two orthogonal axes.  After the ML prediction, we rotated back the MTPs in the original, global frame.

All MTP interactions were computed in CHARMM 
\cite{brooks2009charmm} using the MTPL module \cite{bereau2013scoring, 
bereau2013leveraging}, while our in-house code used for the many-body 
dispersion energies \cite{bereau2014toward} is freely available online \cite{mbvdw}.

\subsection{Voronoi partitioning of the charge density}
\label{sec:vor}

\begin{figure}[htbp]
  \begin{center}
    \includegraphics[width=0.8\linewidth]{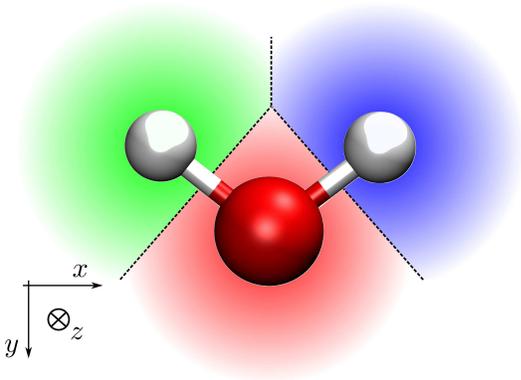}
    \caption{Cartoon of Voronoi-based density partitioning for a
      water molecule (rendered with VMD \cite{humphrey1996vmd}).  Dashed
      lines delineate the partition boundaries.  The axis system illustrates
      the orientation of the water molecule's global frame for the
      calculations presented in Sec.~\ref{sec:wat}: aligned in the $xy$ plane
      with the oxygen atom pointing toward the $y$ axis.}
    \label{fig:voronoi}
  \end{center}
\end{figure}

The Voronoi baseline model relies on a systematic, geometry-dependent estimation of a
system's underlying charge density.  Reference atomic MTP coefficients are extracted
from the partitioning of said density (see below for details), 
where monopole, dipole, and quadrupole
contributions are given by
\begin{align}
  q^{(i)} &= \int \rd \br \, n_i(\br), \label{eq:q} \\
  \mu_\alpha^{(i)} &= \int \rd \br \, n_i(\br) \, r_\alpha, \label{eq:mu}\\
  Q_{\alpha\beta}^{(i)} &= \int \rd \br \, n_i(\br) \, r_\alpha \, r_\beta, \label{eq:quad} 
\end{align}
respectively, where $n_i(\br)$ denotes the \emph{partitioned} density attributed
to atom $i$, as a function of spatial coordinate $\br$, and $\alpha, \beta \in
\{x, y, z\}$.  Rather than being derived from quantum-chemical calculations,
$n_i(\br)$ is constructed as a Gaussian-based atomic density
\begin{equation}
  n_i(\br) = \frac{1}{(2\pi r_i^2)^{3/2}}\exp\left(-\frac{|\br - {\bf R}_i|^2}{2
    r_i^2}\right),
\end{equation}
where ${\bf R}_i$ is the position of nucleus $i$, and $r_i$ is the chemical
element's free-atom radius which is fixed independent of molecular environment
or geometry. For this, we have used parameters reported
elsewhere~\cite{Chu-Dalgarno, anatole-jcp2010}.  Atomic densities
$\{n_i(\br)\}$ are partitioned according to a Voronoi scheme
\cite{okabe2009spatial}, whereby only the closest atom contributes to a given
spatial coordinate.  The Euclidean distance provides the distance metric to
identify a region $\mathcal{R}_p$ associated to atom $p$
\begin{equation}
  \label{eq:voronoi}
  \mathcal{R}_p = \{{\bf r} \in \mathbb{R}^3 \;|\; d({\bf r}, {\bf r}_p) \leq
  d({\bf r}, {\bf r}_j)\; \text{for all}\; j \neq p\}.
\end{equation}
Fig.~\ref{fig:voronoi} illustrates the Voronoi-based density partitioning
between the atoms of water. Each color corresponds to the atomic density of the
corresponding atom.  We recently introduced a similar protocol to 
effectively estimate atomic polarizabilities which serve as
input for many-body dispersion interactions
\cite{bereau2014toward}.

Note that the Voronoi model contains no free parameter---the free-atom radii being
applied without prior fitting.  Though the model hardly reproduces any of the
reference MTP coefficients, it provides a \emph{qualitative} evaluation of
the coefficients.  In particular, the baseline model reproduces elementary
symmetries of the system that are entirely determined by the geometry, e.g., a
planar molecule cannot generate an orthogonal dipole moment.

While we compute Eqns.~\ref{eq:q}, \ref{eq:mu}, and \ref{eq:quad} in Cartesian
coordinates, we subsequently convert them to their spherical counterparts
$Q_{\kappa m}$, where $\kappa$ denotes the rank (e.g., $\kappa=0$ for monopoles)
and $m$ indexes the (real) component of the MTP (see Stone
\cite{stone2013theory} for more details).  Given a molecular structure, we
estimate for each atom the baseline property $p^{\rm Vor} = \{ Q_{00}, Q_{1m},
Q_{2m}\}$, where $m$ runs over the $2\kappa + 1$ elements of an MTP of
order $\kappa$.



\subsection{Molecular dataset}

To refine atomic properties beyond the baseline prediction, we train the
transferable ML algorithm on $2,896$ neutral molecules obtained from the
Ligand.Info database \cite{grotthuss2004ligand}, totaling 82.1 kilo atoms.
Atoms have been segregated between training and prediction pools randomly.  The
database provides three-dimensional coordinates of small, organic molecules.
We focus exclusively on a subset of neutral molecules that include elements
H, C, N, O, S, F, and Cl.

\section{Results}

\subsection{Voronoi-based baseline evaluation}
\label{sec:wat}

To illustrate the applicability of the Voronoi-based baseline evaluation of
MTP coefficients, we compare its prediction with the reference MTP
coefficients obtained from {\em ab initio} methods.  Given that MTPs are inherently
axis-system dependent (apart from the monopole), we first describe the global
frame used for the water molecule in Fig.~\ref{fig:voronoi}.
Inherent symmetries of the geometry impose some coefficients to be zero, 
e.g., there can be no dipole moment along the $x$ or $z$ directions due to the molecule's $C_{2v}$ symmetry. 
While the Voronoi-baseline does not even qualitatively reproduce  the
non-zero coefficients---due to the method not entailing {\em any} prior
parametrization---its ability to impose the right symmetry is very desirable. 
The same kind of behavior is also shown for a carbon atom on benzene, or the carbonyl
oxygen of formic acid in Tab.~\ref{tab:coeffs}.
For comparison, Tab.~\ref{tab:coeffs} also shows already the corresponding ML augmented
MTP result.
As such, the baseline recovers an important aspect of the underlying symmetry, 
which the augmenting ML model no longer needs to account for.

\begin{table}[htbp]
  \begin{center}
    \begin{tabular*}
      {\linewidth}{@{\extracolsep{\fill}} r|c|ccc|ccccc}
      & $Q_{00}$ & $Q_{10}$ & $Q_{11c}$& $Q_{11s}$& $Q_{20}$ & $Q_{21c}$&
      $Q_{21s}$& $Q_{22c}$& $Q_{22s}$\\
      \hline
      \hline
      \multicolumn{10}{c}{water oxygen}\\ 
      Vor & 0.04 & 0 & 0 & -0.12 & 0.43 & 0 & 0 & -0.21 & 0.01\\
      $\Delta$-ML&-0.13&-0.01&-0.02&0&0.16&-0.03&0.04&-0.18&0\\
      Ref & -0.39& 0 & 0.01 & -0.40 & -0.92 & 0 & 0 & 0.45 & 0\\
      \hline
      \multicolumn{10}{c}{formic acid carbonyl oxygen}\\ 
      Vor & 0.01 &0& 0& 0& 0.08& 0 & 0 & 0 & 0\\
      $\Delta$-ML&-0.35& -0.30&-0.03& 0.03& 0.55& 0.10&-0.04&-0.10&0.15\\
      Ref        &-0.45& -0.10&-0.15& 0   & 0.38& 0.13& 0 &-0.32 & 0\\
      \hline
      \multicolumn{10}{c}{benzene carbon}\\ 
      Vor & 0.01 &-0.01&0.01& 0 &-0.01 & 0 & 0 & 0.07 & 0.01\\
      $\Delta$-ML&-0.10&-0.04&-0.10&-0.09&-0.73&-0.24&0.19&0.17&-0.05\\
      Ref & -0.03& 0 & 0 & 0 & -0.65 & 0 & 0 & -0.14 & 0\\
      \hline
      \hline
    \end{tabular*}
    \caption{MTP coefficients of oxygen in water 
    (see Fig.~\ref{fig:voronoi}) and in the carbonyl bond of formic acid,
     as well as the MTPs of carbon in benzene.  ``Vor,'' ``$\Delta$-ML,'' and ``ref'' 
    correspond to the Voronoi-based property evaluation, the delta-learned prediction
    across chemical compound space (see the $\Delta$-ML-85 model below), and the ab initio
    data, respectively.  All coefficients are expressed in units $e$\AA$^l$,
    where $l$ is the MTP's rank. The comparatively low accuracy of 
    the $\Delta$-ML model
    for water is rationalized in the main text. 
    }
    \label{tab:coeffs}
  \end{center}
\end{table}

\subsection{$\Delta$-ML MTP model trained and tested across chemical space}
\label{sec:vpml}

In principle, the above-mentioned Coulomb matrix encodes enough chemistry to train all
chemical elements.  Memory limitations of the kernel-ridge regression, 
however, make atom-type specific models better tractable.
We now investigate the $\Delta$-ML model's capabilities to
predict MTP coefficients \emph{across} chemical space, one for
each of those chemical elements that are most frequent in small, organic molecules (see above).
The ML model has been trained on various fractions of the considered dataset's 82
kilo atoms.

\begin{figure*}[htbp]
  \begin{center}
    \includegraphics[width=0.9\linewidth]{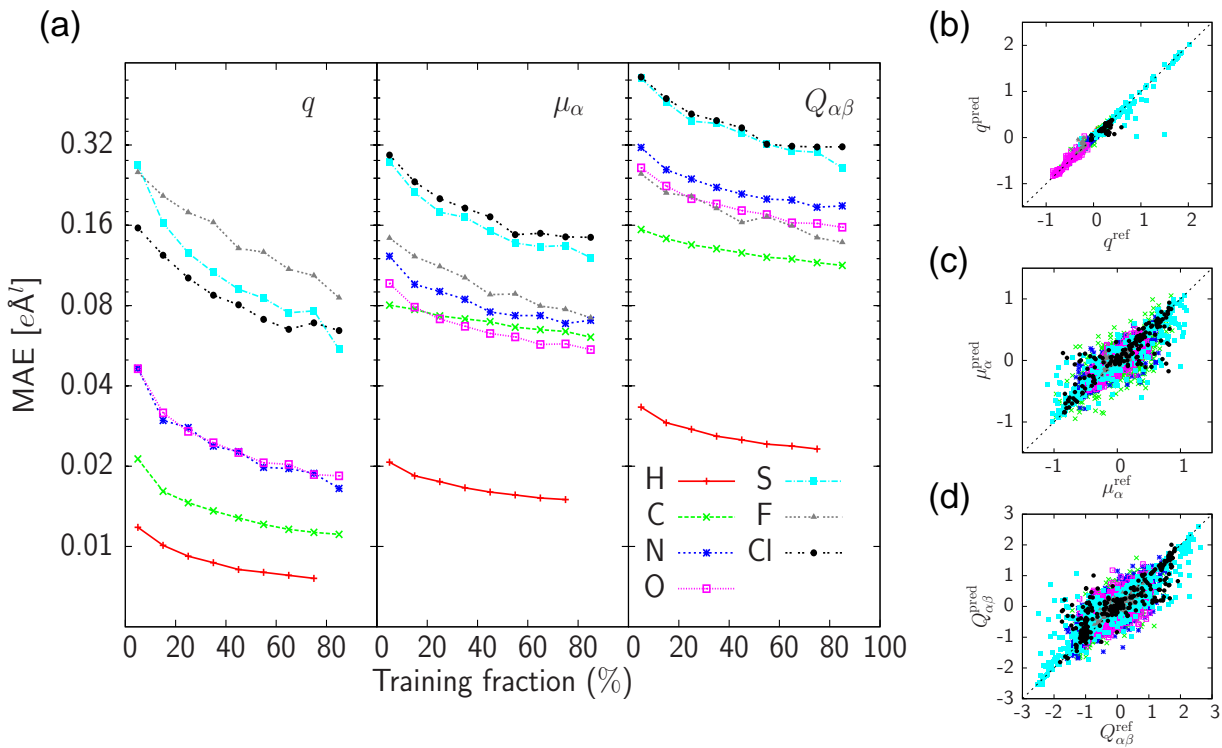}
    \caption{(a) Mean absolute prediction error (MAE) as a function of training set percentage of  an 82k-atom database obtained from neutral molecules. MAEs are given in $e$\AA$^l$ units with $l$ being the MTP rank.
      Errors correspond to MTP $\Delta$-ML model predictions for individual components of atomic 
      monopole, dipole, and quadrupole moments
      shown in left, mid, and right-hand side panel, respectively, for all elements present. 
      Scatter correlation plots for all components of (b) monopoles, (c) dipoles,
      and (d) quadrupoles, as predicted with the $\Delta$-ML-85 model. Colors
      correspond to the atom legend in (a). 
      The outliers in the monopole correlation curve correspond to sulfur-oxide 
      groups (see main text).
    }
    \label{fig:sat_tpt}
  \end{center}
\end{figure*}

Fig.~\ref{fig:sat_tpt} (a) displays error saturation curves for individual chemical elements.  
These monotonically decaying learning curves are presented as a function of training
size of the dataset, where the predicted mean absolute error (MAE) is
calculated across the \emph{remaining} atoms not included in the training set.
The finding of monotonically decreasing error with training set size
represents numerical evidence for a crucial feature of the supervised-learning
working hypothesis: The accuracy of the $\Delta$-ML model of MTPs improves as more data is being added.
Monopole coefficients have the fastest learning rate, quickly 
reaching MAE between $0.1~e$ and $0.01~e$ for all elements. 
Differences between elements are simply due to their relative frequency in the data base. 
More specifically, since hydrogens and carbons predominate in small organic molecules
they provide much larger training sets, and consequently more accurate ML models when measured
in terms of percentage of training data being used. 
The scatter correlation plot between predicted and
reference monopoles for each element is given in Fig.~\ref{fig:sat_tpt}
(b) for the largest training fraction used: 85\% (denoted ``$\Delta$-ML-85''),
the exception being the hydrogen $\Delta$-ML model trained on only 75\% of the dataset
given its converged accuracy.
The monopoles modeled by $\Delta$-ML-85 reach high Pearson correlation coefficients: $R^2 \approx 97\%$, except
F and Cl, for which $R^2 = 17\%$ and $68\%$, respectively. 
Such poor performance is explained by the small size training data set available for these elements. 
The outliers in the monopole scatter plot (Fig.~\ref{fig:sat_tpt} (b)) 
correspond to sulfur-oxide groups. 
Also here, the few samples of these groups in the training set
results in significant prediction errors.

Predicted dipoles show a MAE across elements between $0.02~e$\AA\xspace and $0.15~e$\AA, depending on 
training set size.  
The heterogeneity of the chemical environments of the 
elements is reflected in the ML-model's performance. 
The $\Delta$-ML dipole moments of hydrogens are extremely 
accurate---most likely due to hydrogens showing weak overall MTP 
moments and due to their repeating saturating bonding pattern. 
By contrast, carbon atoms, albeit being nearly as frequent as hydrogens in the
database, have MTP ML models with significantly larger MAE.  
We rationalize the $\Delta$-ML's relative difficulty to predict this element by the large chemical
variety carbon exhibits, i.e., strongly varying hybridization states and possible bonding with all other
elements.  
Also note the reversal of the relative offset of the F and Cl learning curves as one
proceeds from monopole to dipole moments, despite the fact that there are roughly twice
as many Cl as F atoms in the data base. 
This effect is possibly due to chlorine's larger polarizability, which implies that
the chemical environment of the atom plays a more important role for the dipole-moment, 
turning the ML-based modeling into a higher-dimensional and thereby more challenging 
statistical-learning problem. 
Such effects, however, can only fully be explained through an in-depth study with 
significantly larger data sets. 
The scatter plot for predicted versus reference dipole moments is shown  for all elements
in Fig.~\ref{fig:sat_tpt} (c) for $\Delta$-ML-85. 
Clearly, the correlation is worse in comparison to monopoles ($R^2 \approx 50\%$),
which is in line with what one would expect for a more complex vectorial quantity.

Of all MTPs considered, quadrupoles represent the most complex and challenging property. 
Not surprisingly, the resulting ML models yield the largest MAE for our training set:
between $0.02~e$\AA$^2$\xspace and $0.30~e$\AA$^2$ depending on training set size. 
The spread of MAEs across elements is strikingly more pronounced.  
Nevertheless, we find a larger correlation coefficient compared to
dipoles: $R^2 \approx 65\%$, see Fig.~\ref{fig:sat_tpt} (c) and (d).
Note that the Cl/F accuracy reversal with respect to the monopole model is also manifested for 
the $\Delta$-ML MTP model of this rank.

\subsection{$\Delta$-ML MTP vs ML MTP model}

\begin{table*}[htbp]
  \begin{center}
    \begin{tabular*}
      {\linewidth}{@{\extracolsep{\fill}} rrr|cc|cc|cc}
      & & & \multicolumn{6}{c}{MAE [$e$\AA$^l$]} \\
      & \multicolumn{2}{c|}{\# atoms} & \multicolumn{2}{c}{$q$} & 
      \multicolumn{2}{c}{$\mu_\alpha$} & \multicolumn{2}{c}{$Q_{\alpha\beta}$}\\
      & training ($N_t$) & prediction & ML-85 & $\Delta$-ML-85 & ML-85 & 
      $\Delta$-ML-85 & ML-85 & $\Delta$-ML-85 \\
      \hline
      H & 28,822& 9,607 & 0.01 & 0.01 & 0.03 & 0.01 & 0.04 & 0.02 \\
      C & 24,356& 4,297 & 0.01 & 0.01 & 0.05 & 0.05 & 0.18 & 0.09 \\
      N &  4,054&   715 & 0.02 & 0.02 & 0.09 & 0.05 & 0.26 & 0.15 \\
      O &  6,134& 1,082 & 0.02 & 0.02 & 0.08 & 0.04 & 0.22 & 0.12 \\
      F &    363&    63 & 0.03 & 0.09 & 0.04 & 0.05 & 0.14 & 0.11 \\
      S &  1,542&   272 & 0.05 & 0.05 & 0.12 & 0.09 & 0.31 & 0.20 \\
      Cl&    739&   130 & 0.03 & 0.06 & 0.11 & 0.10 & 0.28 & 0.26 \\
      \hline
        & 66,010&16,166 & 0.02 & 0.04 & 0.08 & 0.06 & 0.20 & 0.13 \\
    \end{tabular*}
    \caption{MAE for each chemical element and MTP rank of the ML-85 and
      $\Delta$-ML-85 transferable models (neutral molecules), corresponding
      to an 85\% training-set size for all elements but H (only 75\%).
      The last lines averages over all chemical elements.
      The second column denotes the number of molecules for which the 
      MTP moments have been predicted.
      }
    \label{tab:wV75}
  \end{center}
\end{table*}

We have compared the relative improvement gained when combining the 
ML with the baseline evaluation from the Voronoi scheme.  
Tab.~\ref{tab:wV75} compares
MAEs decomposed by chemical element for the prediction with 85$\%$ training
fraction both with (i.e., ``$\Delta$-ML-85'') and without (i.e., ``ML-85'') prior
Voronoi baseline evaluation.  The table also specifies the number of atoms involved in
the prediction pool (i.e., outside the training fraction).  
While the Voronoi scheme does nothing to improve monopoles (for F and Cl it even worsens
the prediction) it is increasingly helpful as we move to dipoles (with negligible change for F and Cl),
and quadrupoles (with small improvement for F and Cl).
We stress that the observed trends for F and Cl should be interpreted with utmost caution 
since their frequency in the database is very small (363 and 739). 
The lack of improvement for monopoles stems directly from the Voronoi scheme's strategy: 
Merely encoding symmetries, only higher MTP moments can benefit from the absence of a
number of components that are forbidden by the underlying geometry.  For fixed
training size, the MAE is roughly halved for quadrupoles when using the delta
learning procedure, compared to the standard ML methodology.  

All results discussed so far refer to ML models of atomic MTPs in neutral molecules. 
For positively and negatively charged compounds we have
found ML models to yield very similar trends and accuracy (data not shown).

\section{Validation}

To assess the introduced MTP model's applicability we have
used predicted electrostatic coefficients to evaluate intermolecular interaction energies in molecular dimers and organic crystals. 
To do this, we accounted only for static MTP electrostatics and many-body dispersion (MBD) interactions,
\begin{equation}
E_{\rm vdW} \approx E_{\rm MTP} + E_{\rm MBD},
\end{equation}
neglecting induction, penetration, and repulsion terms.
A short description of the MBD formalism is provided in the next paragraph.
As discussed above, our MTP and MBD-models also represent approximations in the form of the $\Delta$-ML model
and dipole-dipole-manybody expansion, respectively.
To better gauge the effect of the introduced MTP-ML model, we
also compare to vdW energy predictions using quantum-mechanically (QM) derived MTPs.

Common approximations in the exchange-correlation potential used in
density functional theory lead to inadequate predictions of dispersion
interactions. This has motivated the development of a number of 
dispersion-corrected methods.  We hereby rely on the method developed
by Tkatchenko and coworkers \cite{MBD}, in which free-atom polarizabilities
are first scaled according to their close environment following a
partitioning of the electron density.  The many-body dispersion up to
infinite order (in the dipole approximation) is then obtained by 
diagonalizing the Hamiltonian of a system of coupled quantum 
harmonic oscillators, thereby coupling the scaled atomic polarizabilities
at long range.  The importance of many-body effects and accuracy of the
method has been demonstrated on a large variety of systems \cite{MBD}.
Later, a classical approximation relaxed
the requirement for an electron density, using instead a physics-based
approach to estimate how atomic polarizabilities needed to be scaled
based on a Voronoi partitioning \cite{bereau2014toward}.

\subsection{Molecular dimers}

\begin{figure}[htbp]
  \begin{center}
    \includegraphics[width=0.9\linewidth]{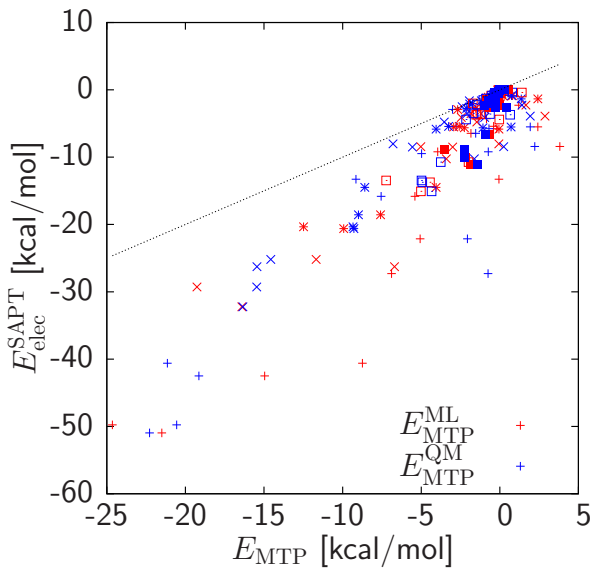}
    \caption{Scatter plot between reference energy components 
    from SAPT and computed electrostatics from either quantum-mechanical
    coefficients, $E_{\rm MTP}^{\rm QM}$, or the universal 
    $\Delta$-ML-85 MTP model, $E_{\rm MTP}^{\rm ML}$, for intermolecular dimers in the
    S22x5 dataset~\cite{grafova2010comparative}. Dimers closer than the equilibrium distance (i.e., factor 0.9)
    were systematically excluded. Symbols correspond to different distance
    factors that multiply the equilibrium distance: 0.9 (plus sign), 1.0 (cross), 1.2 (star), 1.5 (open square), and 2.0 (filled square). 
    The straight line corresponds to ideal correlation.}
    \label{fig:s22x5-sapt}
  \end{center}
\end{figure}

To gauge the accuracy of the electrostatics alone, we compare 
the electrostatic componenent of reference symmetry adapted perturbation theory (SAPT)
results \cite{hesselmann2011comparison, lao2014symmetry} to the corresponding 
intermolecular components derived either from QM MTPs or from the $\Delta$-ML-MTP model.
Fig.~\ref{fig:s22x5-sapt} displays the correlation plot between the two model MTP
electrostatics calculations and SAPT for the S22 dimers\cite{jurevcka2006benchmark}
at different intermolecular distances 
\cite{grafova2010comparative,hesselmann2011comparison, lao2014symmetry}.  The plot
confirms that both MTP models generally underestimate the electrostatic SAPT
component of the interaction energies, presumably due to a lack of 
penetration effects.  Encouragingly, as the intermolecular
distance is increased, the MTP predictions systematically recede to the
perfect correlation line.

\begin{figure*}[htbp]
  \begin{center}
    \includegraphics[width=0.9\linewidth]{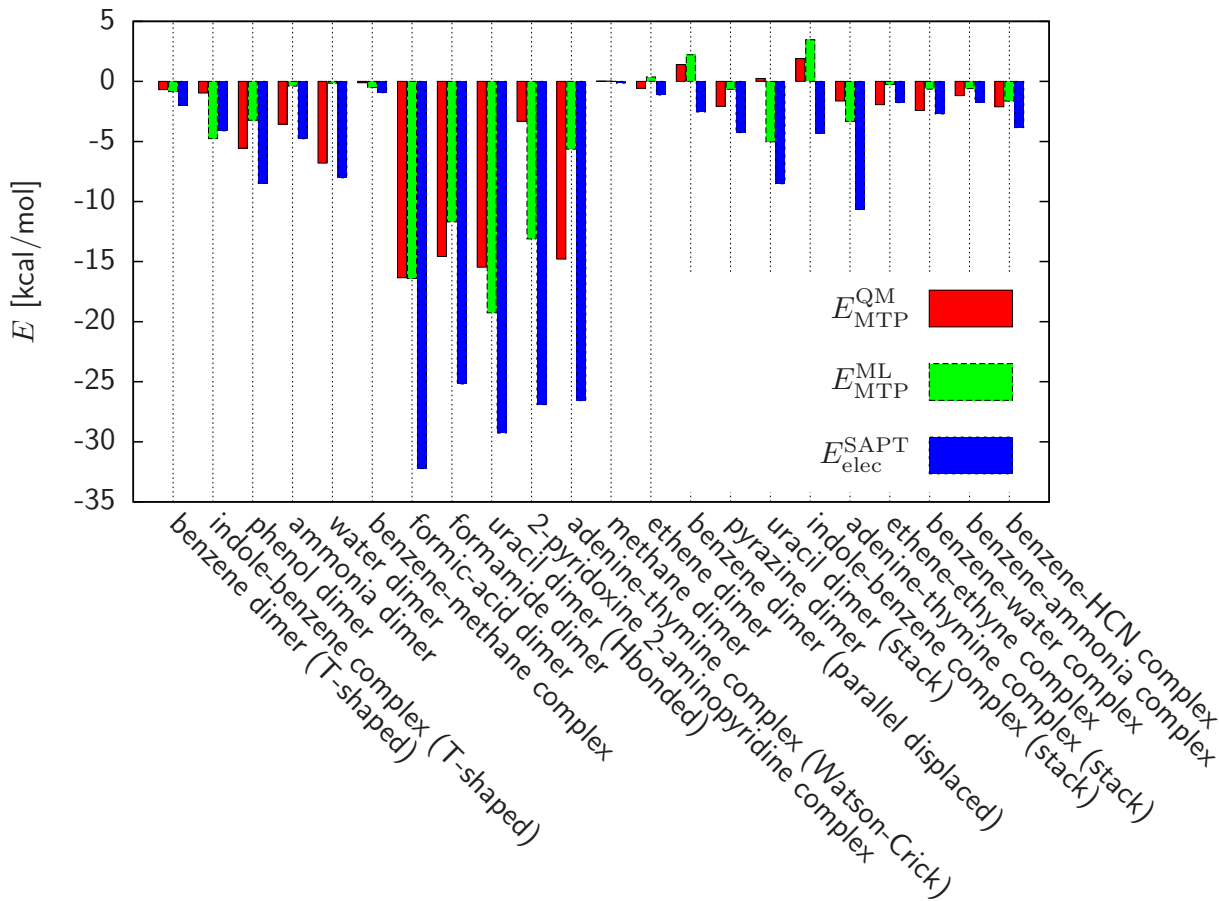}
    \caption{Intermolecular static electrostatic energy contribution for each 
    	dimer of the S22 dataset, for coefficients parametrized from 
        quantum-mechanical calculations, $E_{\rm MTP}^{\rm QM}$, and
        the transferable ML model $\Delta$-ML-85, $E_{\rm MTP}^{\rm ML}$.
        Electrostatic energies from reference SAPT calculations are also
        provided \cite{lao2014symmetry}.}
    \label{fig:s22-decomp}
  \end{center}
\end{figure*}

Not all errors are distributed uniformly across compounds.
Fig.~\ref{fig:s22-decomp} compares the MTP energy contributions $E_{\rm
  MTP}^{\rm QM}$ and $E_{\rm MTP}^{\rm ML}$, as well as reference
SAPT data for each molecular dimer of
the S22 dataset (i.e., at their equilibrium distance). 
We find good correlations between the QM and MTP models, though we note a number of
qualitative discrepancies. In particular, the ML model fails to reproduce the
attractive nature of the electrostatic interaction for the water
and ammonia dimers. Presumably, the model fails to predict their coefficients due to
the molecules' unique chemical composition: the ML model relies on interpolations across
the trained molecules, of which some must be chemically similar to the new
compound.  
Larger molecules are less problematic because similarities between
chemical fragments occur far more frequently. We do find systematic deviations
between the MTP energies and the SAPT reference electrostatic data for
strongly hydrogen-bonding compounds, for which penetration effects
\cite{stone2013theory} become significant.  

We have calculated molecular dimer energies corresponding to various 
datasets for which high-level quantum-chemistry numbers have previously been published. 
We have considered the following databases: S22 \cite{jurevcka2006benchmark}, S22x5 
\cite{grafova2010comparative}, S66 and S66x8 \cite{rezac2011s66}, 
SCAI \cite{berka2009representative}, and X40 and X40x10 
\cite{rezac2012benchmark}. 
All MTP coefficients have been predicted using the $\Delta$-ML-85 model (see 
Fig.~\ref{fig:sat_tpt} and Tab.~\ref{tab:wV75}). 
We only considered dimers made up of the chemical elements H, C, O, N, S, F, and Cl,
keeping 992 out of over 1,300 vdW dimers.

\begin{figure}[htbp]
  \begin{center}
    \includegraphics[width=0.75\linewidth]{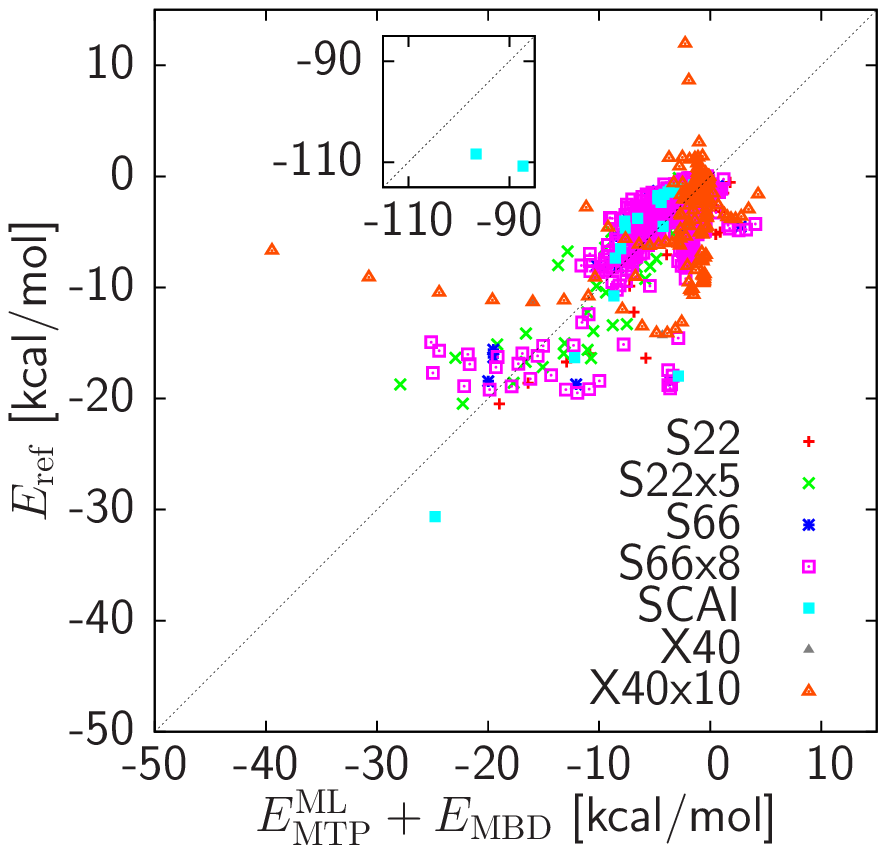}
    \caption{Correlation plot between reference vdW energies, $E_{\rm ref}$, 
    	and energies predicted from the 
    	classical many-body dispersion method with static electrostatics, the latter 
        parametrized from the $\Delta$-ML-85 model, $E_{\rm MTP}^{\rm ML} 
        + E_{\rm MBD}$.  The inset corresponds 
        to the two charged-charged side-chain interactions of the SCAI
        database.  The present lack of repulsion, induction, and penetration
        effects is at cause for the outliers.}
    \label{fig:db}
  \end{center}
\end{figure}

Fig.~\ref{fig:db} contrasts the scatter correlation between reference intermolecular energies, 
$E_{\rm ref}$, and the sum of many-body dispersion and ML-predicted MTP 
electrostatics, $E_{\rm MTP}^{\rm ML} + E_{\rm MBD}$.
The mean-absolute error of all intermolecular estimates using the MTPs from individual quantum-chemistry calculations \cite{bereau2014toward} and the 
ML predictions amount to 2.36 and 2.19~kcal/mol, respectively.
In other words, the ML MTP prediction is on par with MTPs derived from explicit
electron densities generated by computationally demanding quantum-chemistry calculations.
Interactions between charged-charged amino acids of the SCAI database
are reasonably well reproduced (see insets of Fig.~\ref{fig:db}), 
pointing to the robustness of the method not only for neutral compounds, 
but also charged species. 

\subsection{Benzene crystal}

\begin{figure}[htbp]
  \begin{center}
    \includegraphics[width=0.9\linewidth]{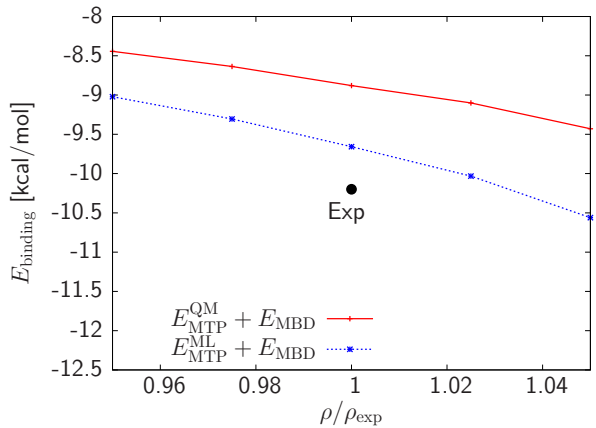}
    \caption{Cohesive binding energy of the benzene crystal as a function of
      the ratio of densities, $\rho/\rho_{\rm exp}$. Both models for
      intermolecular energies are presented: $E_{\rm MTP}^{\rm QM} + E_{\rm
        MBD}$ and $E_{\rm MTP}^{\rm ML} + E_{\rm MBD}$. The experimental value
      is shown explicitly (black dot) \cite{schweizer2006quantum}.}
    \label{fig:bzn}
  \end{center}
\end{figure}

Increasingly accurate and fast
methods provide the means for crystal structure prediction of organic
compounds \cite{bardwell2011towards}, to the point of ranking polymorphs of
molecular crystals \cite{marom2013many, reilly2014role}. 
Moving toward a condensed-phase system, we have also evaluated the 
cohesive binding energy predictions of a molecular benzene crystal.   Following previous
work\cite{meijer1996density, tapavicza2007weakly}, we have computed the
binding energy for different ratios of the unit-cell density with respect to
the experimental density \cite{schweizer2006quantum}, $\rho/\rho_{\rm exp}$.
Isotropic density scalings were performed without further optimization, and
the binding energy included interactions with the neighboring unit cells, as
discussed in our previous publication \cite{bereau2014toward}.
Fig.~\ref{fig:bzn} features the MTP and MBD based vdW estimates of cohesive
energy as a function of density.  Again, we compare the resulting numbers once
using the $\Delta$-ML-predicted MTP electrostatics, combined with MBD, and
once using the QM MTP model combined with MBD.  We find very good agreement
between the two methods, validating here as well the ML model.  Coincidentally,
the two methods yield cohesive binding energies in good agreement with the
experimental value. The lack of repulsive interactions prohibits a
further increase in energy as $\rho$ gets larger \cite{bereau2014toward}. 

\section{Conclusion}
We have introduced machine-learning models for electrostatic multipoles (MTPs) of H, C, O, N, S, F, and Cl atom types. The models have been trained on atomic multipole coefficients of small organic molecules evaluated using first principles calculations. Neutral, cationic, and anionic molecular states were treated with separate models. The model yields highly accurate MTPs for H, reasonable performance for C, N, O, and significant errors for  S, F and Cl due to their sparsity in the training set.

Focusing on the intermolecular S22 dimer dataset, MTP energies show good correlation 
between the coefficients parametrized from ML and individual quantum-chemistry calculations. A comparison with reference electrostatic interactions from symmetry-adapted perturbation theory (SAPT) is satisfactory for large intermolecular separations, and impaired by the lack of penetration effects at short distances.
Furthermore, MTPs from the ML model have been combined with a 
classical many-body dispersion potential to estimate intermolecular
energies of nearly 1,000 molecular dimers as well as the cohesive energy of the
benzene crystal.  The results show that the ML model retains overall a similar 
accuracy compared to calculations with the MTPs parametrized from 
individual quantum-chemical calculations. 

The $\Delta$-ML approach, which augments a physics-based baseline model by a ML model, has proven to be useful to more efficiently train vector and tensor quantities. Incorporation of molecular symmetries via 
the Voronoi partitioning of the charge density, included in the baseline model, is at the heart of this improvement.
 
The proposed models alleviate the need for quantum-chemistry calculations for every single molecule/molecular conformation in a perturbative evaluation of intermolecular interactions, bringing us one step forward toward the task of automated parametrizations of accurate state-independent and transferable force fields. 

\section{Acknowledgments}
We thank Andreas Hesselmann and John M.~Herbert for kindly providing the 
SAPT reference energies, as well as Aoife Fogarty and Markus Meuwly for 
critical reading of the manuscript. 
OAvL acknowledges funding from the Swiss National Science foundation (No. PP00P2\_138932). 
This research used resources of the Argonne Leadership Computing Facility at 
Argonne National Laboratory,
which is supported by the Office of Science of the U.S. DOE under contract DE-AC02-06CH11357. 

\bibliographystyle{ieeetr}
\bibliography{biblio}

\end{document}